# Single Gateway Placement in Wireless Mesh Networks

D.A. Turlykozhayeva *, W. Waldemar , A.B. Akhmetali , N.M. Ussipov ,
S.A. Temesheva , S.N. Akhtanov

[1] Lublin University of Technology,
Faculty of Electrical Engineering and Computer Science,
Department of Electronics and Information Technology, Lublin, Poland
*e-mail:symbat.temesheva@gmail.com

Wireless Mesh Networks (WMNs) remain integral to various sectors due to their adaptability and scalability, providing robust connectivity where traditional wired networks are impractical. From facilitating smart city initiatives to enabling disaster recovery efforts and supporting industrial automation, WMNs play a pivotal role in modern networking applications. Their versatility extends to rural connectivity, underscoring their relevance across diverse scenarios. Recent research in WMNs has honed in on optimizing gateway placement and selection to bolster network performance and ensure efficient data transmission. This paper introduces a novel approach to maximize average throughput by strategically positioning gateways within the mesh topology. Inspired by Coulomb's law, which has been previously employed in network analysis, this approach aims to improve network performance by strategically positioning gateways for optimization. Comprehensive simulations and analyses carried out in this research demonstrate the effectiveness of the proposed method in enhancing both throughput and network efficiency. By leveraging physics-based models such as Coulomb's law, the study offers an objective means to optimize gateway placement—a critical component in WMN design. These findings provide invaluable insights for network designers and operators, guiding informed decision-making processes for gateway deployment across a spectrum of WMN deployments. As such, this research contributes significantly to the ongoing evolution of WMN optimization strategies, reaffirming the indispensable role of gateway placement in establishing resilient and efficient wireless communication infrastructures.

## 1 Introduction

Wireless Mesh Networks (WMNs) continue to hold significance and offer potential in contemporary networking landscapes due to their flexibility, scalability, and capacity for self-organization. They find applicability across a broad spectrum of uses, spanning from household networking to expansive urban installations and industrial environments [1]. WMNs prove especially adept at expanding network reach and capability in a resource-efficient manner. Their adaptability makes them particularly well-suited for scenarios where the installation of conventional wired networks is either economically impractical or logistically challenging [2]. WMNs can dynamically respond to alterations in network structure, environmental factors, and user requirements.

The structure of a Wireless Mesh Network (WMN) involves mesh clients (MCs), gateways (GWs), and mesh routers (MRs). MRs with low mobility form a wireless backbone network within their designated service areas, serving as both traffic relays and access points for MCs [3]. This dual functionality renders them suitable for diverse usage scenarios. GWs hold pivotal roles in network architecture, bridging connectivity between the wireless backbone (comprising MRs) and external networks such as the Internet. They act as ingress and egress points for data traffic entering or exiting the WMN, facilitating essential communication pathways. Previous studies have highlighted the importance of strategically positioning gateways, with a focus on addressing challenges such as identifying the most effective locations and quantities of gateways. Researchers have extensively investigated this aspect [7-11]. For instance, in [12], authors propose heuristic methods utilizing various wireless link models to iteratively select the best gateway positions in order to fulfill Quality of Service (QoS) requirements for associated nodes. In another study [13], an algorithm is introduced to recursively compute the minimum weighted dominating set, aiding in optimal gateway placements while upholding user QoS expectations. It's noteworthy that determining the most suitable gateway placements necessitates intricate computations, thus inviting further exploration [14].

This research proposes enhancing throughput and efficiency in Wireless Mesh Networks by optimizing the placement of a single gateway, leveraging Coulomb's law. This approach draws from a method previously employed by Zhang et al. in the analysis of fractal networks [15].

This article is structured as follows: Section 2 discusses the issue of gateway node placement. Section 3 presents our approach for calculating gateway placement. Finally, Section 4 outlines the key findings and conclusions derived from our investigation.

**2 Gateway node placement**

The Generalized Network Design Problem (GNP) is categorized as NP-hard [16]. Past studies have concentrated on developing efficient heuristic algorithms to achieve near-optimal solutions. Typically, the GNP is formulated as an Integer Linear Program (ILP) to ascertain the minimum number of gateway nodes needed for a given WMN backbone network, while adhering to specific Quality of Service (QoS) constraints. These constraints usually encompass communication delay, relay load, and gateway throughput, all of which significantly influence network performance.

Numerous heuristic algorithms have been proposed to tackle the GNP, aiming to minimize the number of gateway nodes while satisfying the mentioned QoS constraints [16-19]. Additionally, some research has integrated a secondary objective of reducing the number of hops between a mesh router (RN) and its designated gateway [20, 21]. One method suggested for addressing the combined RNP-GNP problem involves employing the heuristic graph clustering technique delineated in [22] among these algorithms.

**3 Model description**

The box-covering method used in fractal network algorithms by Zhang et al. was adopted to calculate the single gateway placement of the network [23]. Two nodes within the network, designated as node 1 and node 2, each characterized by a specific electric charge q1 and q2 and separated by a distance r, were considered. The electrostatic interaction force between these nodes was determined by formula (1):

$$F = k \frac{|q_1 q_2|}{r^2}, \qquad (1)$$

The interaction strength between nodes is directly proportional to the multiplication of their attributes and inversely proportional to the square of the distance between them. Our research methodology involves modelling interactions within complex networks using a node degree-based repulsion algorithm. In this algorithm, we consider each node's degree as its 'charge' in the network and compute repulsive forces based on node connectivity. The first step is to calculate the Euclidean distances between nodes to measure their straight-line distance in a two-dimensional space. Then, we systematically evaluated the interactions between network nodes, including quantifying the cumulative force exerted by neighbouring nodes for each node. Then, we systematically evaluated the interactions between network nodes, including quantifying the cumulative force exerted by neighbouring nodes for each node. Then, we systematically evaluated the interactions between network nodes, including quantifying the cumulative force exerted by neighbouring nodes for each node. This is essential for gauging its overall influence within the network. By using Coulomb's law, we calculated the electrostatic interaction force between nodes within a specific coverage radius. We then accumulated the resulting forces to represent the combined impact of neighbouring nodes on each individual node. The calculated forces were stored for further analysis, which allowed us to identify influential nodes by sorting them based on their computed forces.

The analysis concluded by measuring the average network throughput in NS3, which provided crucial insights into the network's performance. To simulate real-world data flow scenarios, each node was treated as a gateway and linked to a server (NS3) responsible for packet transmission. By systematically connecting a server to each gateway node and calculating the average throughput, we gained a comprehensive understanding of how data transmission efficiency varied across different nodes within the network.

**4 Results**

This section presents the results that demonstrate the impact of gateway location on force and throughput for two types of topologies: grid and random. To construct the networks, we used the NS-3 simulation program with the parameters outlined in Table 1. To determine the throughput between nodes in these networks, we used the existing OLSR routing algorithm.

**Table 1 –** Parameters used for NS-3 simulation program

| Network simulator | NS-3.40 |
| --- | --- |
| Channel type | Wireless channel |
| Propagation model | Friis Propagation Loss Model |
| Network interface type | Phy/Wireless Phy |
| Mac type | Mac/802.11ac |
| Interface Queue type | Drop Tail/PriQueue |
| Link layer type | LL |
| Antenna Model | Single Antenna |
| Traffic type | CBR |
| Transport protocol | UDP |
| Simulation time | 50 s |
| Packet size | 1024 |
| Simulation area | 1000m*1000m |
| Mobility model | Constant |
| Protocol | OLSR |

In Figure 1(a), an example of a deployed grid network configuration of 5x5, consisting of 25 nodes, is shown. In this network, four mesh routers are highlighted with their respective coverage depicted by a yellow circle, and in the middle, a gateway node is specially chosen as a yellow node. Figure 1(b) illustrates the calculated force per throughput ratios in a grid network.

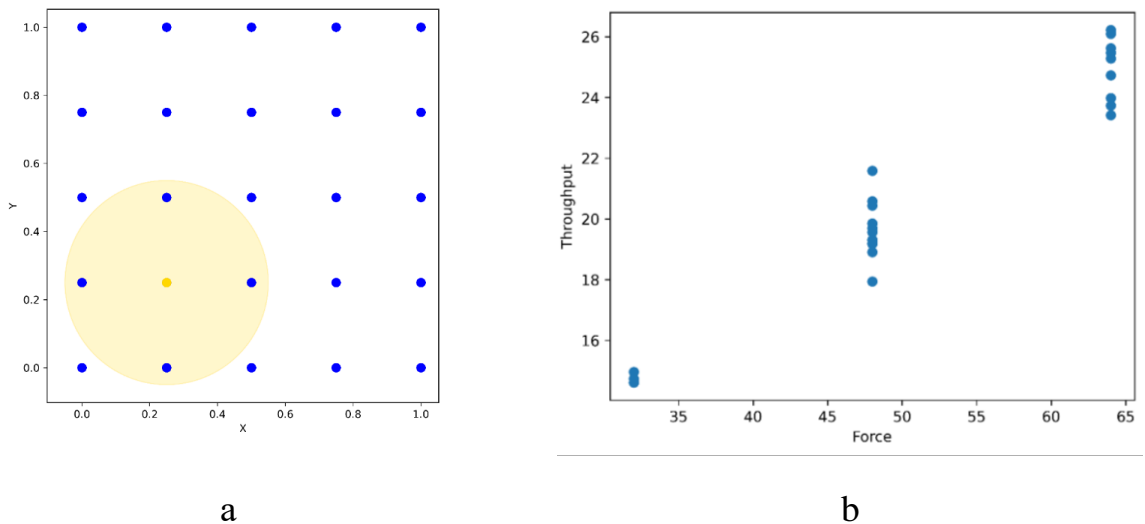

a                                                            b

**Figure 1** (a) – Illustration of a Grid network with a Gateway placed at the Center; (b) – Force-to-Throughput Ratio

Figure 1(b) illustrates a direct positive correlation between force and throughput, suggesting that higher force values correspond to increased throughput. Selecting a node with the highest force could serve as an effective strategy for gateway placement. In Figure 2 (a), a random network comprising 25 nodes is depicted, where 9 mesh routers along with their respective coverage highlighted by a yellow circle and a centrally located gateway node is specially chosen as a yellow node. Figure 2 (b) illustrates the calculated force per throughput ratios in a random network.

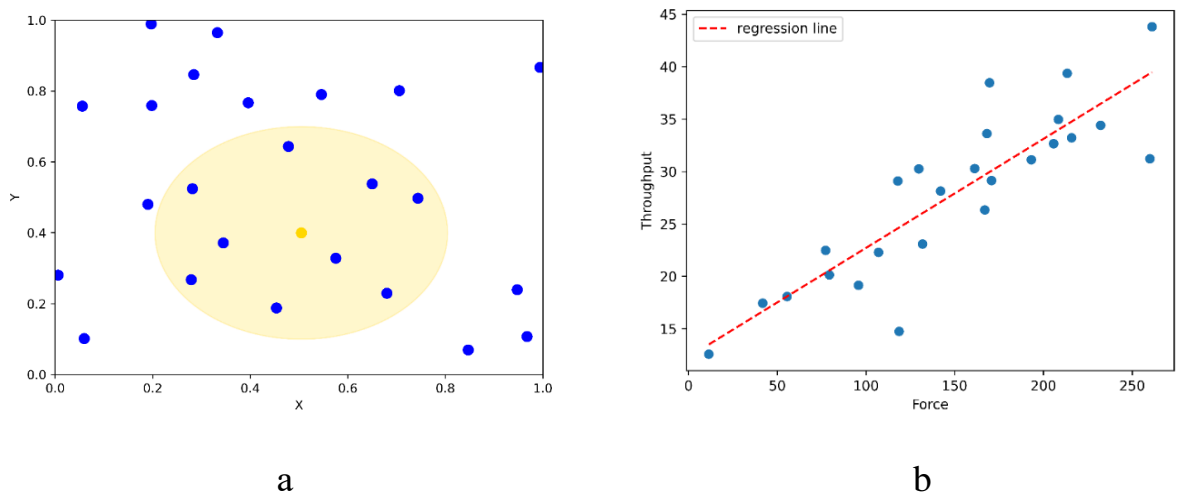

a                                                            b

**Figure 2** (a) – Illustration of a random network with a Gateway placed at the Center; (b) – Force-to-Throughput Ratio

Figure 2 (b) demonstrates a clear direct correlation between force and throughput, indicating that higher force values are associated with increased throughput. Opting for a node with the highest force might prove to be an efficient strategy for selecting gateway placement. These results underscore the critical role of gateway placement in determining network throughput and emphasize the importance of strategic decision-making in network design and optimisation efforts.

**5 Conclusions**
Our study investigated the impact of gateway placement on two common network topologies: random and grid. Our analysis revealed distinct behaviours in each topology.
In the random network topology, which is characterized by a decentralized and irregular node arrangement, strategic gateway placement remained crucial. Although the network is inherently unpredictable, our findings suggest a clear correlation between force and throughput, indicating that optimal gateway positioning can significantly enhance network performance.

In contrast, in the grid network topology, where nodes are organized in a structured and uniform manner, the impact of gateway placement was more pronounced. By strategically placing gateways within the grid, we observed a direct positive correlation between force and throughput. This study emphasizes the significance of taking network topology into account when devising gateway placement strategies. Structured layouts may provide more predictable outcomes.

Our research contributes to the ongoing efforts in optimizing wireless mesh networks and highlights the importance of gateway placement as a key factor in achieving robust and efficient wireless communication networks. Further research could investigate additional factors and algorithms to enhance WMN performance in various deployment scenarios by improving gateway placement strategies.

**Acknowledgments**

We would like to express our sincerest gratitude to the Research Institute of Experimental and Theoretical Physics of the Al-Farabi Kazakh National University for supporting this work by providing computing resources (Department of Physics and Technology). This research was funded by the Ministry of Science and Higher Education of the Republic of Kazakhstan, grant AP19674715 ('Routing of wireless mesh networks based on box-covering algorithms').